\documentclass[prl,aps,twocolumn,superscriptaddress,notitlepage,10pt]{revtex4-2}

\usepackage{printlen}

\usepackage{amssymb,amsfonts,amsmath,amstext,graphicx,hyperref}
\usepackage{amsmath}
\usepackage{tikz}
\usepackage{graphicx}
\usepackage{physics}
\usepackage[normalem]{ulem}
\usepackage{tabularx}

\def\ket#1{|{#1}\rangle}

\def\BraVert{\egroup\,\mid\,\bgroup}

\usepackage{braket}
\usepackage{here}
\hypersetup{
	colorlinks=true,
	linkcolor=blue,
	filecolor=magenta,      
	urlcolor=cyan,
}
\usepackage{xcolor}
\begin{document}

\title{Hierarchical Time Crystals}

\author{Jan Carlo Schumann}
\email{jan.schumann@student.uni-tuebingen.de}
\affiliation{Institut f\"ur Theoretische Physik and Center for Integrated Quantum Science and Technology, Universit\"at T\"ubingen, Auf der Morgenstelle 14, 72076 T\"ubingen, Germany}
\author{Igor Lesanovsky}
\affiliation{Institut f\"ur Theoretische Physik and Center for Integrated Quantum Science and Technology, Universit\"at T\"ubingen, Auf der Morgenstelle 14, 72076 T\"ubingen, Germany}
\affiliation{School of Physics and Astronomy and Centre for the Mathematics and Theoretical Physics of Quantum Non-Equilibrium Systems, University of Nottingham, Nottingham, NG7 2RD, United Kingdom}
\author{Parvinder Solanki}
\email{parvinder@mnf.uni-tuebingen.de}
\affiliation{Institut f\"ur Theoretische Physik and Center for Integrated Quantum Science and Technology, Universit\"at T\"ubingen, Auf der Morgenstelle 14, 72076 T\"ubingen, Germany}
\affiliation{Department of Physics, Indian Institute of Space Science and Technology, Thiruvananthapuram, Kerala 695547, India}
\date{\today}

\begin{abstract}
Spontaneous symmetry breaking is one of the central organizing principles in physics. 
Time crystals have emerged as an exotic phase of matter, spontaneously breaking the time translational symmetry, and are mainly categorized as discrete or continuous.
While these distinct types of time crystals have been extensively explored as standalone systems, intriguing effects can arise from their mutual interaction.
Here, we demonstrate that a time-independent coupled system of discrete and continuous time crystals induces a simultaneous twofold temporal symmetry breaking, resulting in a \textit{hierarchical time crystal} phase.
Interestingly, one of the subsystems breaks a discrete temporal symmetry that is absent from the time-independent Liouvillian generating the dynamics, but instead emerges dynamically, giving rise to a convoluted non-equilibrium phase.
We demonstrate that hierarchical time crystals are robust, emerging for fundamentally different coupling schemes and persisting across wide ranges of system parameters.
\end{abstract}

\date{\today}

\maketitle 
    
\emph{Introduction.---} 
Spontaneous symmetry breaking (SSB), where the stable state of a system lacks a symmetry respected by its dynamical generator~\cite{AnIntroductionToSymmetryBreaking}, shapes phases of matter across scales -- from macroscopic phenomena in cosmology~\cite{symmetry_breaking_cosmology,symmetry_breaking_cosmology2,symmetry_breaking_cosmology3,symmetry_breaking_cosmology4} to the smallest scales of nuclear and particle physics~\cite{symmetry_breaking_nuclear,HiggsSSB,nambu2014spontaneous}.
In many-body quantum systems, SSB has emerged as a unifying framework for organizing both equilibrium and non-equilibrium phenomena \cite{AndersonMoreIsDifferent, sym2020609, morandi_spontaneous_1985, PatternFormationOutofEquilib, ElseDiscreteTimeCrystals}.
Striking manifestations of SSB are time crystal phases, where a many-body system spontaneously breaks the time translational symmetry of the underlying generator of the dynamics~\cite{Wilczek_time_crystals,Bruno_no_go,Watanabe_nogo,Time_crystals_a_review,sacha2020time, khemani2019briefhistorytimecrystals}.
Depending on the nature of the broken temporal symmetry, time crystals can be mainly categorized as discrete (DTC)~\cite{DTC_with_dissipation,Else_FloquetTC, pizzi_higher-order_2021, Prethermal_DTC, DTC_ex_1, DTC_ex_2, DTC_ex_3,RydbergTC,zhu_dicke_2019, subharmonicsOberreiter, FloquetTC_LMG, zhang_observation_DTC_2017,beatrez2023critical,liu_higher-order_2024,ShuoDTC,choi_observation_2017} or continuous (CTC)~\cite{iemini2018btc, System_with_feedback, ExactSolutionBTC, Seeding_crystallization,PhysRevLett.130.150401, BTC_ex_1, BTC_ex_2, ExoticSynchronization,jirasek2025boundarytimecrystallight,tucker_shattered_2018, CTC_example, huang_observation_2025, Kongkhambut_CTC, liu_bifurcation_2025,solanki2025chaos,solanki2025entanglement,jhd4-1khw,wu_dissipative_2024,Paulino_2026,PhysRevB.106.224308} time crystals. 
A DTC constitutes a phase of matter that breaks a discrete temporal symmetry of its generator, producing subharmonic oscillations that stabilize in the thermodynamic limit.~\cite {sacha2020time,Time_crystals_a_review,Colloquium_DTC}. 
On the other hand, a CTC features persistent oscillations even in the complete absence of time-dependent external fields, breaking the continuous time translational symmetry~\cite{iemini2018btc}.\\
\indent All standalone time crystal setups investigated so far only break a single temporal symmetry.
While the interplay between multiple, nested symmetries has recently garnered interest in many-body systems  \cite{nandkishore_fractons_2019,FractonPhasesOfMatter,HigherFormSubsysSymmetryBreaking}, its implications for temporal symmetries remain limited to time-dependent closed quantum systems \cite{EngineerHierarchicalSymm}.\\
\begin{figure}[t!]
\centering
\includegraphics{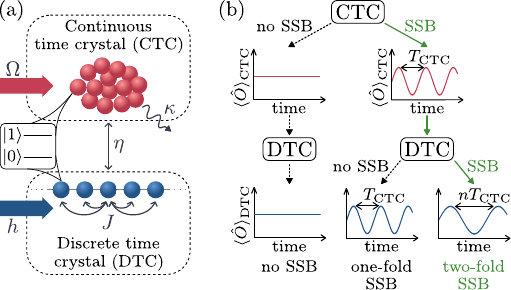}
\caption{\textbf{Sketch of the model and hierarchical symmetry breaking.} 
(a) The time-independent coupled time crystal model features a CTC subsystem, realized as a collective spin ensemble, which interacts with a DTC subsystem of long-range coupled spins at interaction strength $\eta$. Each spin-$1/2$ system consists of a ground state $\ket{0}$ and an excited state $\ket{1}$.
Both CTC and DTC are subjected to constant drives with strengths $\Omega$ and $h$, respectively, with the CTC additionally exhibiting collective emission at a rate $\kappa$.
(b) Illustration of spontaneous symmetry breaking (SSB) possibilities: no SSB, where the observables $\langle \hat{O} \rangle_\alpha$ are time-independent for both CTC and DTC; one-fold SSB, where only the CTC breaks a temporal symmetry, leading to synchronized periodic oscillations of both DTC and CTC with period $T_{\text{CTC}}$; and twofold SSB, where the DTC discretely breaks the emergent periodicity of the CTC phase, leading to subharmonic oscillations with a period $T_{\text{DTC}} = nT_{\text{CTC}}$. The latter phase is termed a hierarchical time crystal (HTC).
}
\label{fig: HSB explanation}
\end{figure}
    In this work, we demonstrate that distinct temporal symmetries can be broken simultaneously in a time-independent system, giving rise to a convoluted phase of matter, which we term a \textit{hierarchical time crystal} (HTC). To uncover this intriguing phenomenon, we consider a coupled system comprising a CTC and a DTC (displayed in Fig.~\ref{fig: HSB explanation}(a)).
    Here, the underlying hierarchical breaking of multiple time translational symmetries unfolds in two steps, as illustrated in Fig.~\ref{fig: HSB explanation}(b). 
    First, the CTC can spontaneously break the continuous time translational symmetry of the time-independent composite system, leading to an emergent oscillatory phase. 
    The hereby emerging time-periodicity of the CTC can then, in turn, be broken discretely by the DTC, manifesting in a subharmonic response to the CTC phase (green path in Fig.~\ref{fig: HSB explanation}(b)). 
    Interestingly, the DTC breaks a symmetry that is absent in the underlying dynamical generator, but rather emerges dynamically, thereby establishing an intricate non-equilibrium phase of matter characterized by nested temporal order. 
    Remarkably, the DTC subsystem subharmonically locks to the CTC subsystem, exhibiting both integer and fractional multiples of the CTC period.
    This locking phenomenon does not require fine-tuning; instead, it features multiple plateaus of distinct locking frequencies spanning over a range of system parameters, resulting in a staircase-like structure associated with higher-order synchronization phenomena \cite{SynchronStaircase, SynchronStaircase2, wu_farey_2022, feng2025tunablediscretequasitimecrystal}.
    Crucially, this staircase is not dictated by an external periodic drive, as the reference periodicity is itself spontaneously generated.
    We demonstrate that the emerging HTC phase is robust against parameter variations, quantum fluctuations, and fundamentally different coupling mechanisms between the time crystals, encompassing both coherent and dissipative interactions.
    The associated concept of hierarchical symmetry breaking extends beyond merely temporal symmetries, introducing a broader framework of hierarchical dissipative phase transitions.

\emph{HTC model.---} 
    To investigate the emergence of an HTC, we analyze a coupled system of a CTC and a DTC, governed by the Lindblad master equation
       \begin{equation}
        \dot{\hat{\rho}}=\mathcal{L}_\text{HTC}\hat{\rho}=(\mathcal{L}_\text{DTC}+\mathcal{L}_\text{CTC}+\mathcal{L}_\text{Int})\hat{\rho}.
        \label{eq: total Liouvillian}
    \end{equation}
    Here,
    $\mathcal{L}_\text{DTC}$ and $\mathcal{L}_\text{CTC}$ describe the dynamics of the uncoupled DTC and CTC subsystems, respectively, while $\mathcal{L}_\text{Int}$ accounts for their mutual interaction. 
    The combined Liouvillian $\mathcal{L}_\text{HTC}$ entails the dynamics of the composite system.
    
    We now consider specific examples for each subsystem.
    For the CTC, we choose the well-known boundary time crystal model~\cite{iemini2018btc}.
    It consists of a collection of $N_{\text{C}}$ spin-$1/2$ systems that are subject to a constant drive at strength $\Omega$ and experience collective dissipation at rate $\kappa$ (cf. top panel in Fig.~\ref{fig: HSB explanation}(a)). 
    Consequently, the CTC dynamics is governed by the following Lindblad dynamical generator
    \begin{align}
        \mathcal{L}_{\text{CTC}}\hat{\rho} = -i\Omega[\hat{S}_{\text{C}}^x,\hat{\rho}] + \frac{\kappa}{S_{\text{C}}} \mathcal{D}[\hat{S_{\text{C}}^-}]\hat{\rho},
        \label{eq: Liouvillian CTC}
    \end{align}
    where $\mathcal{D}[\hat{O}]\hat{\rho} = \hat{O}\hat{\rho}\hat{O}^{\dagger} - \frac{1}{2} \{\hat{O}^{\dagger}\hat{O}, \hat{\rho}\}$ denotes the dissipator and we set $\hbar=1$.
    The total spin of the CTC subsystem is given by $S_{\text{C}}=N_{\text{C}}/2$, and the collective spin operators are defined as $\hat{S}^{\alpha}_{\text{C}} = \sum_{i=1}^N \hat{\sigma}_{i,\text{C}}^{\alpha}/2$, where $\hat{\sigma}_{i,\text{C}}^{\alpha}$ are the Pauli spin matrices of the $i$-th atom in direction $\alpha\in \{x,y,z \}$. 
    The ladder operators are defined as $\hat{S}_{\text{C}}^{\pm} = \hat{S}_{\text{C}}^x \pm i\hat{S}_{\text{C}}^y$. 
    In the thermodynamic limit $N \to \infty$, the system transitions from a stationary state in the strongly dissipative regime ($\Omega/\kappa<1$) to an oscillatory phase in the weakly dissipative regime ($\Omega/\kappa>1$), breaking the generator's continuous time translational symmetry. 
    This dissipative phase transition manifests in the closing of the Liouvillian spectral gap~\cite{MingantiSpectralTheory} as the system size diverges, with the emergence of purely imaginary eigenvalues for the time crystal phase ~\cite{iemini2018btc, MuhleBTCTrajectories, ExactSolutionBTC}.
    
    For the DTC subsystem, we choose a variant of the Lipkin–Meshkov–Glick (LMG) model \cite{LMG1,LMG2,LMG3}, which is known to develop a robust subharmonic response under periodic driving~\cite{pizzi_higher-order_2021, FloquetTC_LMG}. 
    In particular, it consists of a chain of $N_{\text{D}}$ spin-$1/2$ systems that are connected via long-range interactions at strength $J$ (cf. Fig.~\ref{fig: HSB explanation}(a)).
     Here, we consider the case of a static drive at strength $h$, where the system's dynamics can be described by a Lindblad generator of the form 
    \begin{align}
        \mathcal{L}_{\text{DTC}}\hat{\rho} = -i \left[-\frac{2J}{S_{\text{D}}}(\hat{S}_{\text{D}}^z)^2 - h\hat{S}_{\text{D}}^x , \hat{\rho}\right]. 
        \label{eq: Liouvillian DTC}
    \end{align}
    The collective DTC spin operators, $S^{x,y,z,\pm}_{\text{D}}$, are defined analogously to the CTC case, where the subscript D (C) refers to the DTC (CTC) subsystem.
    For simplicity, we assume in the following that both subsystems contain the same number of constituents $N$, such that $S_{\text{C}}=S_{\text{D}}=N/2$.
    After detailing the specific time-crystal models, we explore how various coupling schemes influence the emergence of HTCs in the following sections.

\emph{Coherent coupling.---}
    \begin{figure*}[]
        \centering
        \includegraphics{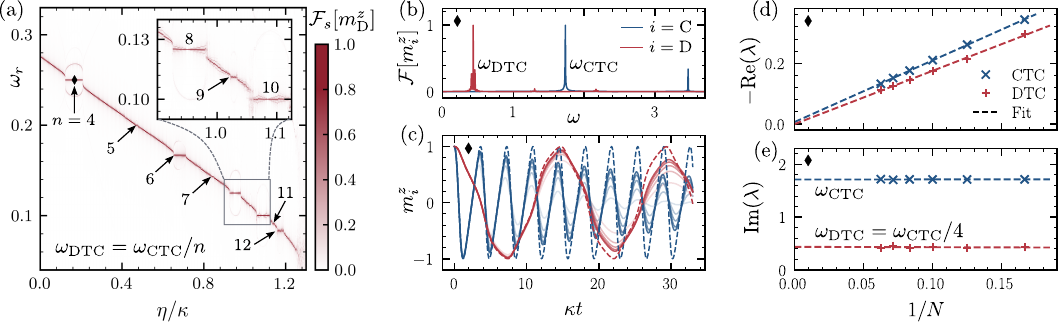}
        \caption{\textbf{Hierarchical symmetry breaking via coherent coupling.} In panel (a), stroboscopic Fourier amplitudes of the DTC magnetization $z$ component $\mathcal{F}_s[m_{\text{D}}^z]$ are plotted against the rescaled inter-TC coupling strength $\eta/\kappa$. 
        The relative frequency $\omega_r=\omega_{\text{DTC}}/\omega_{\text{CTC}}$ exhibits distinct locking plateaus at integer values for the DTC order $n$, where $\omega_{\text{DTC}}$($\omega_{\text{CTC}}$) represents the dominant frequency of DTC (CTC). System parameters are fixed to $\Omega/\kappa=2$, $J/\kappa=0.1$, and $h/\kappa=0.25$, while the initial state is $(m^x_\text{C,D},m^y_\text{C,D},m^z_\text{C,D})=(0,0,1)$. The main panel (inset) probes $m_\text{D}^z$ over 1000 (3000) CTC periods. 
        The HTC phase for the specific case of a $4 $-DTC (diamond in (a)) is detailed in panels (b-e).  
        Panel (b) presents the normalized Fourier spectrum $\mathcal{F}[m_{i}^z]$ of both CTC (blue) and DTC (red). The respective dominant frequencies are labeled. 
        In panel (c), the time evolution of $m_{\text{C}}^z$ (blue) and $m_{\text{D}}^z$ (red) is displayed for finite system sizes $N\in \{ 20,40,\dots,100\} $ with ascending opacity, which approaches the mean-field solutions indicated with dashed lines. 
       Panels (d) and (e) show the dependence of the real and imaginary parts of the dominant Liouvillian eigenvalues on the system size. The dashed lines indicate linear fits to the CTC and DTC eigenvalues, confirming the spectral gap closure with purely imaginary eigenvalues that match the frequencies obtained from the mean-field analysis in panel (b).
     }
        \label{fig: Master figure YX coupling}
    \end{figure*}
    We begin by analyzing a coherent coupling scheme, $\mathcal{L}_{\text{Int}}\hat{\rho}=-i[\hat{H}_{\text{Int}}, \hat{\rho}]$,
    where 
    \begin{align}
        \hat{H}_{\text{Int}} = \frac{\eta}{2iN}\bigg(\underbrace{\hat{S}_{\text{C}}^+\hat{S}_{\text{D}}^- - \hat{S}_{\text{C}}^-\hat{S}_{\text{D}}^+}_{\text{co-rotating}} + \underbrace{\hat{S}_{\text{C}}^+\hat{S}_{\text{D}}^+ - \hat{S}_{\text{C}}^-\hat{S}_{\text{D}}^-}_{\text{counter-rotating}}  \bigg).
    \end{align}
    This interaction Hamiltonian establishes a bilinear coupling between two collective spins, allowing for both excitation exchange (co-rotating terms) and simultaneous excitation creation/annihilation (counter-rotating terms).
    To analyze how this interaction term leads to the hierarchical breaking of temporal symmetries, we first investigate the mean-field dynamics in the thermodynamic limit $N \to \infty$. 
    In particular, we study the time evolution of the expectation values of the rescaled collective spin operators $m_i^{\alpha} \equiv \langle\hat{m}_i^{\alpha}\rangle=2\langle \hat{S}_i^{\alpha}\rangle/N$, assuming the factorization of expectation values due to vanishing correlations such that $\langle \hat{m}_i^\alpha \hat{m}_j^\beta \rangle=\langle \hat{m}_i^\alpha \rangle \langle \hat{m}_j^\beta \rangle$.\\
    \indent To observe HTCs, we consider the weak dissipation regime ($\Omega/\kappa>1$) of the CTC, corresponding to its symmetry-broken time crystal phase where the observables $m_{\text{C}}^{x,y,z}$ begin to oscillate with a finite frequency $\omega_\text{CTC}$ (cf. Fig.~\ref{fig: HSB explanation}(b)) \cite{iemini2018btc}.
    In this setting, the CTC essentially functions as an emergent periodic drive for the DTC, and the coupling parameter $\eta$ takes on the role of the corresponding effective drive strength.
    In contrast to conventional DTC setups, this reference periodicity is generated internally and is bidirectionally coupled to the DTC, allowing mutual backaction.
The DTC then further breaks the emergent periodicity and develops subharmonic oscillations.
    To characterize the associated HTC phase, we consider the ratio of the oscillation frequencies $\omega_r=\omega_\text{DTC}/\omega_\text{CTC}$ as an order parameter, since it captures whether the DTC subharmonically locks to the CTC phase. 
    Here, $\omega_\text{DTC}$ denotes the dominant DTC frequency.\\
    \begin{figure*}[]
        \centering
        \includegraphics{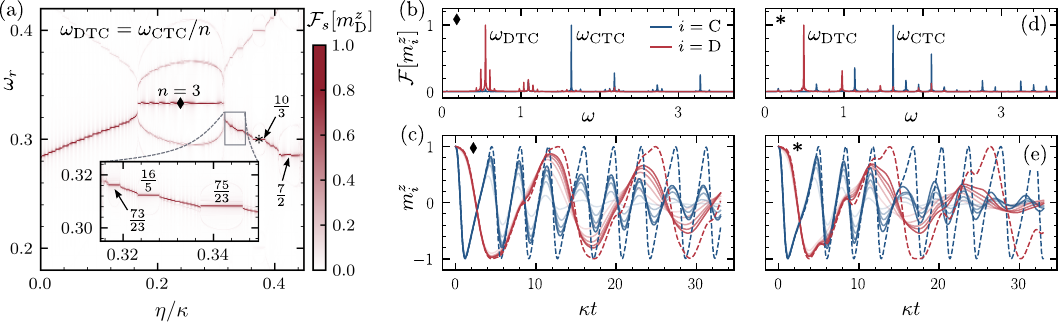}
        \caption{\textbf{Hierarchical symmetry breaking via dissipative coupling.} (a) Stroboscopic Fourier amplitudes of the DTC magnetization $z$ component $\mathcal{F}_s[m_{\text{D}}^z]$ are plotted against the rescaled inter-TC coupling strength $\eta/\kappa$. The dominant component of the relative frequency $\omega_r$ exhibits distinct locking plateaus at both integer and fractional values for the DTC order $n$. System parameters are fixed to $\Omega/\kappa=2$, $J/\kappa=0.08$, and $h/\kappa=0.2522$, while the initial state is $(m^x_\text{C,D},m^y_\text{C,D},m^z_\text{C,D})=(0,0,1)$. The main panel and inset probe $m_\text{D}^z$ over 1000 and 3000 CTC periods, respectively. The system parameters are chosen to target a (b,c) $3$-DTC (diamond in (a)) and a (d,e) fractional $10/3$-DTC (star in (a)), respectively.
        (b),(d) Normalized Fourier spectra of both CTC (blue) and DTC (red) with labeled dominant frequencies.  (c),(e) Time evolution of $m_{\text{C}}^z$ (blue) and $m_{\text{D}}^z$ (red) for finite system sizes $N\in \{ 20,40,\dots,100\}$ displayed in ascending opacity, approaching the dashed lines indicating the corresponding mean-field solution.}
        \label{fig: Master figure dissipative coupling}
    \end{figure*}
    \indent Simulating the mean-field equations~\cite{supp} over a range of coupling strengths $\eta$ reveals a staircase-like structure in the order parameter $\omega_r$ (see Fig.~\ref{fig: Master figure YX coupling}(a)). 
    It features distinct plateaus of constant DTC frequency with respect to the CTC frequency such that $\omega_\text{DTC}=\omega_\text{CTC}/n$.
    The corresponding order $n$ of the $n$-DTC (black in Fig.~\ref{fig: Master figure YX coupling}(a)) prominently exhibits integer values in the range $n=4-12$.
    These stable plateaus indicate robust subharmonic frequency locking behavior, thereby facilitating the emergence of an HTC. We confirm that they persist over a broad range of DTC interaction and drive parameters $J$ and $h$~\cite{supp}, demonstrating the robustness of the HTC phase.
    We illustrate this locking mechanism in Fig.~\ref{fig: Master figure YX coupling}(b), where we analyze the normalized Fourier amplitudes at a stable plateau corresponding to the $4$-DTC (diamond marker in Fig.~\ref{fig: Master figure YX coupling}(a)).
    One observes sharp dominant peaks confirming the subharmonic relation, while occurring secondary peaks corresponding to higher harmonics and indicating non-linear oscillations are significantly smaller, resulting in subtle aperiodic signatures atop the subharmonic response.\\
    \indent To incorporate finite-size effects and intrinsic quantum fluctuations, we go beyond the mean-field limit to finite system sizes $N$. 
    Remarkably, the observed HTC phase remains robust to inter-system backaction and intrinsic quantum fluctuations.
    To reach this insight, we numerically solve Eq.~(\ref{eq: total Liouvillian}) for different $N$. 
    Note that for the calculations, we can restrict ourselves to the symmetric spin-$N/2$ (Dicke) subspace, since the system's total spin is conserved. 
    Additionally, we choose the system parameters such that we operate in a previously found stable $n$-DTC regime, namely $n=4$ (cf. diamond marker in Fig.~\ref{fig: Master figure YX coupling}(a)). 
    The resulting magnetization trajectories for both CTC (blue) and DTC (red) in Fig.~\ref{fig: Master figure YX coupling}(c) clearly exhibit the predicted fourfold increase of the DTC period with respect to the CTC for all system sizes.
    Furthermore, the individual trajectories decay exponentially with a system-size-dependent rate but converge towards the mean-field solutions (corresponding dashed lines) as the system size diverges, a hallmark for the closing of the Liouvillian spectral gap.\\
    \indent To showcase the associated dissipative phase transition, we thus analyze the eigenspectra of $\mathcal{L}_\text{HTC}$ in Eq.~(\ref{eq: total Liouvillian}) for finite system sizes. 
    In particular, we focus on the dominant eigenvalues of the CTC and DTC subsystems responsible for the observed dynamics in Fig.~\ref{fig: Master figure YX coupling}(c).
    They can be distinguished by the imaginary and real parts matching the trajectories' oscillation frequency and exponential decay rate, respectively. 
    The real parts decay as a power law and extrapolate to zero (Fig.~\ref{fig: Master figure YX coupling}(d)), while the imaginary parts displayed in Fig.~\ref{fig: Master figure YX coupling}(e) converge to finite values matching the dominant frequencies of the subsystems from Fig.~\ref{fig: Master figure YX coupling}(b).
    We emphasize the simultaneous gap closing of the subsystems with the associated simultaneous breaking of temporal symmetries, while preserving the fixed frequency ratio across all system sizes $N$. This underlines the stability of the HTC phase, which persists even in the presence of intrinsic quantum fluctuations.
    Furthermore, the emergence of stable HTC phases is not confined to this specific type of coherent coupling, as Ref.~\cite{supp} shows for coherent spin-exchange coupling.

\emph{Dissipative coupling and fractional HTC.---}
The HTC phase emerges independently of the nature of the coupling.
To showcase this, we consider collective dissipation as an incoherent coupling scheme, characterized by the Liouvillian,
 \begin{align}
        \mathcal{L}_{\text{Int}}\hat{\rho} = \frac{2\eta}{N} \mathcal{D}[\hat{S}_{\text{C}}^- + \hat{S}_{\text{D}}^-]\hat{\rho}.
        \label{eq: Interaction Liouvillian Dissipative}
    \end{align}
    Analogous to the preceding section, we begin by analyzing the emerging mean-field dynamics while fixing the CTC in its symmetry-broken phase.
    Similar to the coherent coupling, one observes constant plateaus for the order parameter $\omega_r$ that span over finite ranges of the coupling strength $\eta$, as shown in Fig.~\ref{fig: Master figure dissipative coupling}(a),  
    and persist under variations of the remaining DTC parameters~\cite{supp}, confirming the stability of the emergent DTC phase.
    
    However, a key qualitative difference compared to coherent coupling lies in the emergence of fractional $n$-DTC phases, defined by the relation $n=p/q$, where $p$ and $q$ are coprime integers. 
    This behavior originates from the fact that the distinct coupling schemes induce different nonlinear structures in the mean-field dynamics (see \cite{supp}), thereby reshaping the higher-order frequency-locking landscape of the coupled system.
    To understand this new feature, we compare the Fourier spectra of integer ($3$-DTC) and fractional ($10/3$-DTC) DTC phases, highlighted by the diamond and star markers, respectively, in Fig.~\ref{fig: Master figure dissipative coupling}(a).
    The corresponding normalized spectra are displayed in Figs.~\ref{fig: Master figure dissipative coupling}(b) and (d). 
    Analogous to the discussion of coherent coupling, the spectra for integer locking reveal sharp peaks at dominant frequencies for both subsystems.
    This results in nicely behaved oscillations for both DTC and CTC as observed in Fig.~\ref{fig: Master figure dissipative coupling}(c).
    Conversely, for fractional locking, more secondary peaks with significant magnitude coexist alongside the dominant peaks, rendering highly non-linear oscillations, as depicted in Fig.~\ref{fig: Master figure dissipative coupling}(e).
    This richness in dynamics stems from the higher-order synchronization feedback between subsystems with fractional frequencies, resulting in intricate magnetization trajectories.\\
    \indent A similar effect can be observed even for finite system size trajectories for both fractional and integer cases, which approach the mean-field solutions as the system size diverges.
    Therefore, we confirm the emergence of HTC for dissipative coupling and the validity of mean-field results by performing a finite-size analysis of the exact master equation.
    This further underlines that the emergence of HTC phases is not limited to a particular interaction scheme, but is rather inevitable, even for incoherent collective dissipative coupling.\\
\indent    \emph{Experimental implementation.---}  The intriguing phenomenon of hierarchical symmetry breaking can be explored in existing cavity QED setups that have successfully realized the effective LMG Hamiltonian \cite{sauerwein2023engineering} used for the DTC model and the open Dicke model \cite{sciadv.adu5799}, which leads to a CTC phase.
    The coupling between the CTC and DTC subsystems can be implemented either via cavity-mediated coherent exchange interactions in a shared or coupled cavity geometry, or via collective dissipation into a common lossy photonic mode \cite{Mivehvar02012021, Seeding_crystallization}, as derived in detail in Ref.~\cite{supp}.

   Exciton-polariton condensates provide a complementary and highly versatile platform, where tunable coherent, dissipative, and nonlinear interactions have already enabled the realization of time crystal phases \cite{autti2021ac, doi:10.1126/science.adn7087, PhysRevLett.120.215301}. Moreover, recent work has demonstrated that coupled exciton-polariton condensates can sustain a time-crystalline phase through coherent tunneling and dissipative reservoir coupling \cite{fainstein2026synchronization}, establishing exciton-polariton architectures as a promising setting for the experimental realization of HTC phases. Interacting Bose–Einstein condensate systems coupled to optical cavities constitute another platform in which both DTC \cite{KongkhmabutObservation2} and CTC phases \cite{Kongkhambut_CTC} have been experimentally observed, paving the way for the observation of nested temporal order and hierarchical dissipative phase transitions.

\indent\emph{Conclusion.---} Our findings unveil a mechanism for inducing nested temporal order in non-equilibrium many-body systems.
    In particular, we identified and characterized hierarchical time crystals (HTC) and the underlying hierarchical breaking of temporal symmetries in an interacting open quantum system composed of two distinct types of time crystals: continuous and discrete.
    We demonstrated that the continuous-time translational symmetry broken by the CTC gives rise to an emergent period that is itself discretely broken by the DTC, resulting in twofold symmetry breaking and, consequently, a stable HTC phase.
    Moreover, we showed that HTCs remain robust against parameter variations, quantum fluctuations, and fundamentally different coupling schemes, including both dissipative and coherent mechanisms.
    The HTC thus realizes an autonomous hierarchy of temporal symmetry breaking, in which the nested periodicities provide multiple, independently addressable timescales: a resource directly applicable to time-crystal-based time-keeping~\cite{AutonomousClocks, TCClock}, robust information storage~\cite{esencan2026timecrystalspassivelyprotected}, and metrology~\cite{montenegro_quantum_2023, Albert_sensing}.
    Finally, our results establish HTCs as a robust and broadly realizable organizing principle in driven-dissipative quantum matter, opening a route toward new classes of nonequilibrium phases with nested symmetry breaking. Identifying other platforms capable of nested temporal symmetry breaking remains an open question for future work.

\emph{Acknowledgments.---} P.S. acknowledges support from the Alexander von Humboldt Foundation through a Humboldt research fellowship for postdoctoral researchers. I.L. acknowledges support from the QuantERA II programme (project CoQuaDis, DFG Grant No. 532763411) that has received funding from the EU H2020 research and innovation programme under GA No. 101017733. I.L. also acknowledges funding from the Deutsche Forschungsgemeinschaft (DFG, German Research Foundation) through the Research Unit FOR 5413/1, Grant No.~465199066, from the Leverhulme
Trust (Grant No. RPG-2024-112) and from the European Union through the ERC grant OPEN-2QS (Grant No. 101164443). J.S. acknowledges support from the Deutschlandstipendium for committed students.

\emph{Data availability.---} The data that support the findings of this article are 
available from the authors upon reasonable request.

\onecolumngrid
\newpage

\setcounter{equation}{0}
\setcounter{page}{1}

\setcounter{figure}{0}
\setcounter{table}{0}
\makeatletter
\renewcommand{\theequation}{S\arabic{equation}}
\renewcommand{\thefigure}{S\arabic{figure}}
\renewcommand{\thetable}{S\arabic{table}}

\makeatletter
\renewcommand{\theHequation}{S\arabic{equation}}
\renewcommand{\theHfigure}{S\arabic{figure}}
\renewcommand{\theHtable}{S\arabic{table}}
\makeatother

\setcounter{secnumdepth}{1}

\begin{center}
{\Large SUPPLEMENTAL MATERIAL}
\end{center}
\begin{center}
\vspace{0.8cm}
{\Large Hierarchical time crystals}
\end{center}
\begin{center}
Jan Carlo Schumann,$^{1}$ Igor Lesanovsky,$^{1,2}$ and Parvinder Solanki$^{1}$
\end{center}
\begin{center}
$^1${\em Institut f\"ur Theoretische Physik and Center for Integrated Quantum Science and Technology,}\\
{\em Auf der Morgenstelle 14, 72076 T\"ubingen, Germany}\\
$^2${\em School of Physics and Astronomy and Centre for the Mathematics}\\
{\em and Theoretical Physics of Quantum Non-Equilibrium Systems,}\\ 
{\em The University of Nottingham, Nottingham, NG7 2RD, United Kingdom}
\end{center}
    \noindent In this Supplemental Material, we present detailed mean-field calculations and stability analyses for the coherent and incoherent coupling schemes described in the main text. Additionally, we extend our analysis by discussing a different coherent coupling scheme and provide detailed derivations for the experimental realization of the proposed settings.
\date{\today}

\maketitle 
\appendix
\section{Mean-field equations and stability regimes for coherent coupling}
\label{sec: coherent coupling}
    In this section, we present a detailed discussion of the mean-field results for the coherent coupling scheme outlined in the main text. We also analyze the stability regimes of the hierarchical time crystal (HTC) phase with respect to various system parameters, and discuss the role of the different terms in the system Hamiltonians.\\

    \subsection{Mean-field equations}
    The time evolution of an observable $\hat{O}$ can be given by the following equation
    \begin{equation}
        \begin{aligned}
            \frac{d}{dt}\langle \hat{O}\rangle &= \frac{d}{dt} \Tr\left( \hat{\rho} \hat{O} \right)=\Tr(\mathcal{L}^\dagger\hat{O})\\
            &=i\Tr\left([\hat{H},\hat{O}]\hat{\rho}\right) + \sum_i \frac{\gamma_i}{2}\Tr\left( ([\hat{L}_i^\dagger,\hat{O}]\hat{L}_i + \hat{L}_i^\dagger [\hat{O},\hat{L}_i])\hat{\rho}\right),
            \label{eq: appendix general expect val}
        \end{aligned}
    \end{equation}
    where we inserted the Lindblad master equation in the second step and used the cyclic property of the trace to rearrange the terms. Here, $\hat{H}$ denotes the system Hamiltonian, $\hat{L}_i$ the $i$-th jump operator and $\gamma_i$ the corresponding decay rate. For the considered coherent coupling scheme, the total Hamiltonian reads 
    \begin{align}
    \label{supp: full Hamiltonian YX coupling}
        \hat{H} = \Omega \hat{S}_{\text{C}}^x-\frac{4J}{N}(\hat{S}_{\text{D}}^z)^2 - h\hat{S}_{\text{D}}^x + \underbrace{\frac{\eta}{2iN}\bigg(\hat{S}_{\text{C}}^+\hat{S}_{\text{D}}^+ - \hat{S}_{\text{C}}^-\hat{S}_{\text{D}}^- + \hat{S}_{\text{C}}^+\hat{S}_{\text{D}}^- - \hat{S}_{\text{C}}^-\hat{S}_{\text{D}}^+ \bigg)}_{\hat{H}_\text{Int}}.
    \end{align}
   Note that by plugging in the definition of the spin raising and lowering operators $\hat{S}_j^\pm = \hat{S}_j^x \pm i \hat{S}_j^y$, the expression for the interaction Hamiltonian can be simplified to $\hat{H}_\text{Int}=2\eta  \hat{S}_{\text{C}}^y \hat{S}_{\text{D}}^x/N$. In the considered setting, we only have one jump operator $\hat{L}_1 = \hat{S}_{\text{C}}^-$ with corresponding decay rate $\gamma_1=2\kappa/N$.
    We now consider $\hat{O} = \hat{m_i}^{\alpha}$ with the rescaled collective spin operators $\hat{m}_i^{\alpha}=2\hat{S}_i^{\alpha}/N$ introduced in the main text, where $\alpha \in \{x,y,z\}$ and $i\in \{\text{C,D}\}$.
    In the thermodynamic limit $N\to \infty$, these operators commute, since
    \begin{align}
        \norm{[\hat{m}_j^{\alpha},\hat{m}_k^{\beta}]} = \frac{\delta_{jk}}{N}i\epsilon^{\alpha \beta \gamma}\norm{\hat{m}_j^{\gamma}} \xrightarrow{N \to \infty} 0.
    \end{align}
    We assume vanishing correlations, defining the mean-field limit together with $N\to\infty$, such that
    \begin{align}
        \langle \hat{m}_j^\alpha \hat{m}_k^\beta \rangle=\langle \hat{m}_j^\alpha \rangle \langle \hat{m}_k^\beta \rangle.
    \end{align}

    Calculating Eq.~(\ref{eq: appendix general expect val}) for the rescaled collective spin operators using that their expectation values factor in the mean-field limit yields a set of coupled nonlinear differential equations for their expectation values $m_i^{\alpha} \equiv \langle \hat{m}_i^{\alpha}\rangle$. These mean-field equations define the dynamics of the system in the thermodynamic limit. For the specified coherent coupling scheme, one obtains:
    \begin{equation}
        \begin{aligned}
            \frac{d}{dt}m_{\text{C}}^x &= \eta m_{\text{C}}^z m_{\text{D}}^x + \kappa m_{\text{C}}^z m_{\text{C}}^x \\
            \frac{d}{dt}m_{\text{C}}^y &= -\Omega m_{\text{C}}^z + \kappa m_{\text{C}}^z m_{\text{C}}^y \\
            \frac{d}{dt}m_{\text{C}}^z &= \Omega m_{\text{C}}^y - \eta  m_{\text{C}}^x m_{\text{D}}^x - \kappa \left((m_{\text{C}}^x)^2 + (m_{\text{C}}^y)^2 \right) \\
            \frac{d}{dt}m_{\text{D}}^x &= 4Jm_{\text{D}}^z m_{\text{D}}^y \\
            \frac{d}{dt}m_{\text{D}}^y &= -\eta m_{\text{C}}^{y} m_{\text{D}}^z + 2hm_{\text{D}}^z - 4Jm_{\text{D}}^z m_{\text{D}}^x\\
            \frac{d}{dt}m_{\text{D}}^z &= \eta m_{\text{C}}^{y} m_{\text{D}}^y -2h m_{\text{D}}^y \;.
        \end{aligned}
    \end{equation}

    \subsection{Stability regimes}
    \label{sec: stability regimes YX coupling}
    \begin{figure}
        \centering
        \includegraphics{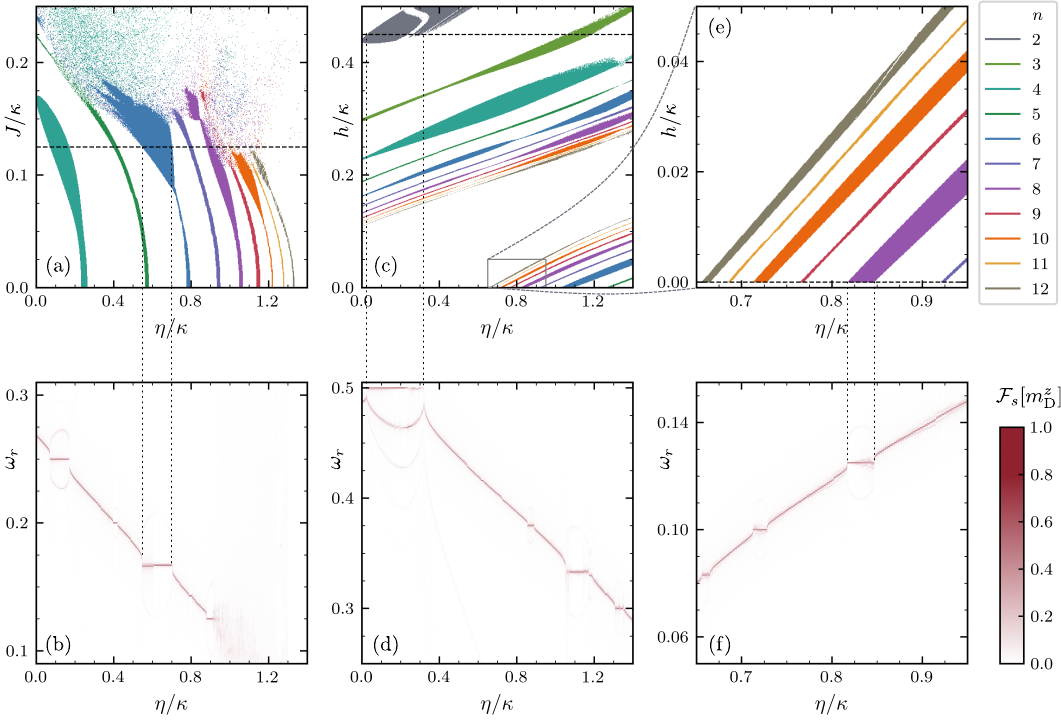}
        \caption{\textbf{HTC parameter regimes for coherent coupling.} (a,c) Dominant relative frequencies of the corresponding $n$-DTC are marked in their respective color. The CTC drive strength is fixed at $\Omega/\kappa=2$, while the initial state is $(m^x_\text{C,D},m^y_\text{C,D},m^z_\text{C,D})=(0,0,1)$. (a) DTC all-to-all interaction strength $J/\kappa$ vs. inter-TC coupling strength $\eta/\kappa$ with $h/\kappa=0.25$. (c) DTC drive strength $h/\kappa$ vs. inter-TC coupling strength $\eta/\kappa$ with $J/\kappa=0.1$. (b,d) Stroboscopic Fourier amplitudes of the DTC magnetization z component $\mathcal{F}_s[m_{\text{D}}^z]$ are plotted against the inter-TC coupling strength $\eta/\kappa$ for a fixed parameter set. The CTC is fixed in its time crystal regime with $\Omega/\kappa=2$ and $m_{\text{D}}^z$ is probed over 1000 CTC periods. (b) Fixed DTC all-to-all interaction strength at $J/\kappa=0.125$ corresponding to the dashed line in (a). Locking intervals of $\eta/\kappa$ coincide with the predicted intervals in (a), as exemplified by grey dotted lines for the $6$-DTC. (d) Fixed DTC drive strength at $h/\kappa=0.45$ corresponding to the dashed line in (c). The predicted locking intervals in (c) are clearly distinguishable (e.g., grey dotted lines for the $3$-DTC). Additionally, a new, fractional locking plateau emerges at $\eta/\kappa \approx 0.92$, which has not been tracked in panel (c). (e,f) Analogous to panels (c,d) but for small values of the DTC drive $h$, demonstrating that the HTC phase persists also for $h=0$ as displayed in (f).}
        \label{fig: Stability regimes y coupling}
    \end{figure}
    In the main text, we demonstrated the emergence of hierarchical symmetry breaking for finite ranges of the inter-TC coupling strength $\eta$, depicted in Fig.~2(a) of the main text. 
    We now investigate the robustness of the associated HTC phase with respect to other parameters of this model, in particular, the all-to-all interaction strength $J$ and the constant drive strength $h$ of the DTC subsystem, and discuss the different roles of the terms in Eq.~(\ref{supp: full Hamiltonian YX coupling}).
    
    To this end, we first examine the stability of the locking plateaus in Fig.~2(a) of the main text as $J$ is varied while $h$ is kept fixed. 
    In particular, we highlight the parameter regimes where we observe stable subharmonic locking of the DTC to the CTC phase.
    This is displayed in Fig.~\ref{fig: Stability regimes y coupling}(a), where the marker color indicates the order $n$ of the corresponding $n$-DTC. 
    Following this logic, uni-colored areas directly correspond to frequency locking, which is stable to perturbations of the system parameters. 
    Indeed, one observes contiguous patches of distinct colors, indicating that the locking behavior is robust to variations in the parameter $J$ and is not fine-tuned to particular values chosen in the main text. 
    
    Furthermore, the staircase structures, such as Fig.~2(a) of the main text, which illustrates the frequency locking steps, are merely a horizontal cut at the corresponding parameter value.
    For the shown cut at $J/\kappa=0.15$ (dashed horizontal line in Fig.~\ref{fig: Stability regimes y coupling}(a)), one receives the stroboscopic Fourier amplitudes displayed in Fig.~\ref{fig: Stability regimes y coupling}(b). Here, the intervals where the DTC sub-harmonically locks to the CTC phase exactly match the predictions from Fig.~\ref{fig: Stability regimes y coupling}(a), for example, illustrated with dotted vertical lines for the $6$-DTC.
    To reobtain Fig.~2(a) in the main text, one performs the horizontal cut at $J/\kappa=0.1$, crossing different stability areas with $n\in\{4,5,\dots,12\}$, manifesting in the discussed rich staircase-like structure.
    Therefore, one can extract the form of the staircase of the relative frequency $\omega_r$ for different parameter values from such stability regime diagrams.
    This also entails predictions for no locking behavior, e.g., for $J/\kappa=0.2$ in Fig.~\ref{fig: Stability regimes y coupling}(a), since a horizontal cut does not cross any stability areas.

    Analogously, in Fig.~\ref{fig: Stability regimes y coupling}(c), we fix the inter-DTC coupling strength $J$ and repeat the above-described procedure while varying the DTC drive strength $h$. Again, we find stable locking regimes that span finite parameter ranges of $h$ and $\eta$, thereby confirming the stable many-body phase. As discussed above, Fig.~2(a) of the main text can be obtained by a horizontal cut at $h/\kappa = 0.25$. 
    Cutting at $h/\kappa=0.45$ yields the Fourier amplitudes displayed in Fig.~\ref{fig: Stability regimes y coupling}(d). As expected, all predicted locking steps from Fig.~\ref{fig: Stability regimes y coupling}(c) can be found. 
    Interestingly, an additional fractional locking step emerges at $\eta/\kappa \approx 0.92$ with $n=8/3$. 
    The corresponding stability region is not shown in Figs.~\ref{fig: Stability regimes y coupling}(c) as it only contains information about integer values of $n$.
    In both panels (a,c) of Fig.~\ref{fig: Stability regimes y coupling}, we find that stability regimes corresponding to higher DTC orders $n$ are generally more sensitive to parameter fluctuations. This can be seen from the decreasing area of the locking regimes for higher $n$ values. 

   Using the insights from above, we are now able to discuss the distinct roles of the different terms in Eq.~(\ref{supp: full Hamiltonian YX coupling}). Apart from the inter-TC coupling $\eta$, which mediates the synchronization between the two subsystems, the DTC dynamics is governed by the all-to-all interaction strength $J$ and the constant drive $h$, which play complementary roles in shaping the HTC phase.
    The drive $h$ sets the bare oscillation frequency of the DTC subsystem in the uncoupled case, and thus determines which subharmonic ratios are accessible for locking. This is reflected in the predominantly horizontal arrangement of the locking plateaus in Fig.~\ref{fig: Stability regimes y coupling}(c): varying $h$ shifts the possible values of $n$ that the DTC locks to, while a fixed $h$ admits stable locking over a finite range of $\eta$. Notably, $h$ is not strictly required for HSB to occur. As shown in Fig.~\ref{fig: Stability regimes y coupling}(e,f), stable subharmonic locking persists even for $h=0$, demonstrating that the drive merely modulates the spectrum of accessible plateaus rather than enabling the synchronization itself.
    
    In contrast, the all-to-all interaction strength $J$ controls the stability of the locking plateaus. For small values of $J$, the locking behavior essentially breaks down, as evidenced by the decreasing width of the colored regions in Fig.~\ref{fig: Stability regimes y coupling}(a). This is consistent with $J$ being the term responsible for the time-crystalline behavior of the bare DTC subsystem, so that a sufficiently strong all-to-all interaction is required for robust locking of the DTC to the CTC.

\newpage
\section{Dissipative coupling}
\label{sec: dissip coupling}
\begin{figure}
        \centering
        \includegraphics{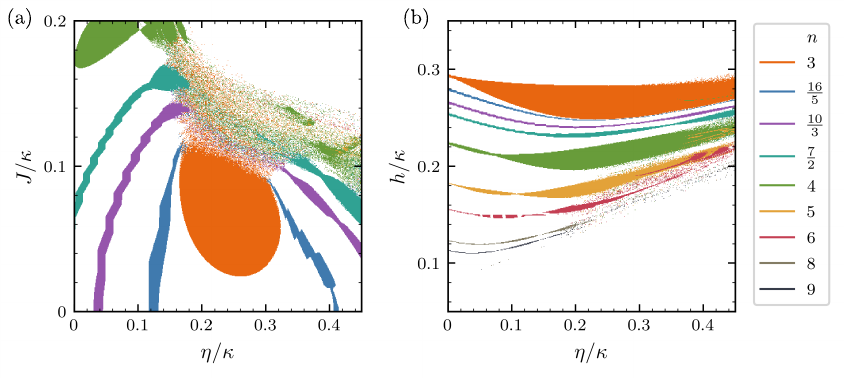}
        \caption{\textbf{HTC parameter regimes for dissipative coupling.} Dominant relative frequencies of the corresponding $n$-DTC are marked in their respective color. The CTC drive strength is fixed at $\Omega/\kappa=2$, while the initial state is $(m^x_\text{C,D},m^y_\text{C,D},m^z_\text{C,D})=(0,0,1)$. (a) DTC all-to-all interaction strength $J/\kappa$ vs. inter-TC coupling strength $\eta/\kappa$. (b) DTC drive strength $h/\kappa$ vs. inter-TC coupling strength $\eta/\kappa$.}
        \label{fig: Stability regimes dissipative coupling}
    \end{figure}
    Analogous to Appendix~\ref{sec: coherent coupling}, this section provides the mean-field equations and HTC stability regimes for dissipative coupling. As described in the main text, the system Hamiltonian in this setting is given as 
    \begin{align}
         \hat{H} = \Omega \hat{S}_{\text{C}}^x-\frac{4J}{N}(\hat{S}_{\text{D}}^z)^2 - h\hat{S}_{\text{D}}^x.
    \end{align}
    As for the previously discussed case of coherent coupling, we consider the jump operator $\hat{L_1} = \hat{S}_{\text{C}}^-$ with decay rate $\gamma_1=2\kappa/N$. Additionally, we include a second jump operator $\hat{L}_2 = \hat{S}_{\text{C}}^- + \hat{S}_{\text{D}}^-$ with decay rate $\gamma_2 = 2\eta/N$, which induces incoherent coupling between the time crystals.
    Plugging this in Eq.~(\ref{eq: appendix general expect val}) for the rescaled collective spin operators $\hat{m}_i^{\alpha}$ as explained in Appendix~\ref{sec: coherent coupling} yields the mean-field equations. One obtains
    \begin{equation}
        \begin{aligned}
            \frac{d}{dt} m_{\text{C}}^x &= \kappa m_{\text{C}}^x m_{\text{C}}^z + \eta m_{\text{C}}^z (m_{\text{C}}^x+m_{\text{D}}^x)\\
            \frac{d}{dt} m_{\text{C}}^y &= \kappa m_{\text{C}}^y m_{\text{C}}^z - \Omega m_{\text{C}}^z + \eta m_{\text{C}}^z (m_{\text{C}}^y +m_{\text{D}}^y)\\
            \frac{d}{dt} m_{\text{C}}^z &= \Omega m_{\text{C}}^y - \kappa \left((m_{\text{C}}^x)^2 + (m_{\text{C}}^y)^2 \right) - \eta \left( m_{\text{C}}^y (m_{\text{C}}^y + m_{\text{D}}^y) + m_{\text{C}}^x(m_{\text{C}}^x + m_{\text{D}}^x)\right)\\
            \frac{d}{dt} m_{\text{D}}^x &= 4J m_{\text{D}}^z m_{\text{D}}^y + \eta m_{\text{D}}^z (m_{\text{C}}^x + m_{\text{D}}^x) \\
            \frac{d}{dt} m_{\text{D}}^y &= \eta m_{\text{D}}^z (m_{\text{C}}^y + m_{\text{D}}^y) + 2h m_{\text{D}}^z -4J m_{\text{D}}^z m_{\text{D}}^x \\
            \frac{d}{dt} m_{\text{D}}^z &= -2h m_{\text{D}}^y - \eta (m_{\text{D}}^y(m_{\text{C}}^y + m_{\text{D}}^y) + m_{\text{D}}^x (m_{\text{C}}^x + m_{\text{D}}^x)).
        \end{aligned}
    \end{equation}
    To investigate the HTC stability regimes, we mark parameter sets that correspond to stable subharmonic frequency locking of the DTC with respect to the CTC. The result is displayed in Fig.~\ref{fig: Stability regimes dissipative coupling}, where we investigate the dependence on the inter-DTC interaction strength $J$ and the DTC drive strength $h$ in panels (a) and (b), respectively. 
    As discussed above, the locking step behavior displayed in Fig.~3(a) in the main text can be obtained by a horizontal cut at $J/\kappa=0.08$ in Fig.~\ref{fig: Stability regimes dissipative coupling}(a), or equivalently at $h/\kappa=0.2522$ in Fig.~\ref{fig: Stability regimes dissipative coupling}(b).
    As expected, one finds continuous areas of one color in both panels of Fig.~\ref{fig: Stability regimes dissipative coupling}, most prominently for the $3$-DTC (orange) and $4$-DTC (green), which indicate stable $n$-DTC, and therefore HTC phases. 
    In Fig.~\ref{fig: Stability regimes dissipative coupling}(a), stable areas of locking to fractional multiples of the CTC period are also clearly distinguishable (blue and purple). The corresponding fractional DTC phases are, however, notably more sensitive to variations of $h$ as evident from Fig.~\ref{fig: Stability regimes dissipative coupling}(b).
    Interestingly, in both subplots, a seemingly larger area of stable locking is intersected by thin strips where the locking breaks down, forming, for example, a honeycomb-like structure.
    Nevertheless, we confirm the stability of the subharmonic response across finite ranges of the system parameters, and with that, the stable HTC phase.
    As in the coherent coupling scheme, the all-to-all interaction $J$ predominantly governs the robustness of the locking plateaus, while the drive $h$ selects which subharmonic ratios become accessible, again reflected in the horizontal orientation of the plateaus in Fig.~\ref{fig: Stability regimes dissipative coupling}(b).

\section{Spin-exchange coupling}
\label{supp: sec spin exchange}
    In this section, we discuss an additional coherent coupling scheme, namely spin-exchange coupling. Here, the two subsystems can exchange an excitation coherently.
    Analogous to the discussed coupling schemes in the main text, it is modulated by the inter-TC coupling strength $\eta$.
    The interaction Hamiltonian thus reads
    \begin{align}
        \hat{H}_\text{Int} = \frac{2\eta}{N} \left( \hat{S}_{\text{C}}^+ \hat{S}_{\text{D}}^- + \hat{S}_{\text{C}}^- \hat{S}_{\text{D}}^+\right),
    \end{align}
    which leads to the interaction Liouvillian $\mathcal{L}_\text{Int} = -i [\hat{H}_\text{Int},\hat{\rho}]$.
    We focus on the mean-field limit to get an overview of the system's dynamics. The mean-field equations for the given coupling scheme take the form
    \begin{equation}
        \begin{aligned}
            \frac{d}{dt}m_{\text{C}}^x &= 2\eta m_{\text{C}}^z m_{\text{D}}^y + \kappa m_{\text{C}}^z m_{\text{C}}^x \\
            \frac{d}{dt}m_{\text{C}}^y &= -\Omega m_{\text{C}}^z - 2\eta m_{\text{C}}^zm_{\text{D}}^x + \kappa m_{\text{C}}^z m_{\text{C}}^y\\
            \frac{d}{dt}m_{\text{C}}^z &= \Omega m_{\text{C}}^y + 2\eta (m_{\text{C}}^y m_{\text{D}}^x - m_{\text{C}}^x m_{\text{D}}^y) - \kappa \left((m_{\text{C}}^x)^2 + (m_{\text{C}}^y)^2 \right)  \\
            \frac{d}{dt}m_{\text{D}}^x &= 4Jm_{\text{D}}^z m_{\text{D}}^y + 2\eta m_{\text{D}}^z m_{\text{C}}^y\\
            \frac{d}{dt}m_{\text{D}}^y &= -2\eta m_{\text{D}}^z m_{\text{C}}^x +2hm_{\text{D}}^z -4J m_{\text{D}}^zm_{\text{D}}^x \\
            \frac{d}{dt}m_{\text{D}}^z &= 2\eta (m_{\text{D}}^y m_{\text{C}}^x -m_{\text{D}}^x m_{\text{C}}^y) -2h m_{\text{D}}^y \;.
        \end{aligned}
    \end{equation}

    \begin{figure}
        \centering
        \includegraphics{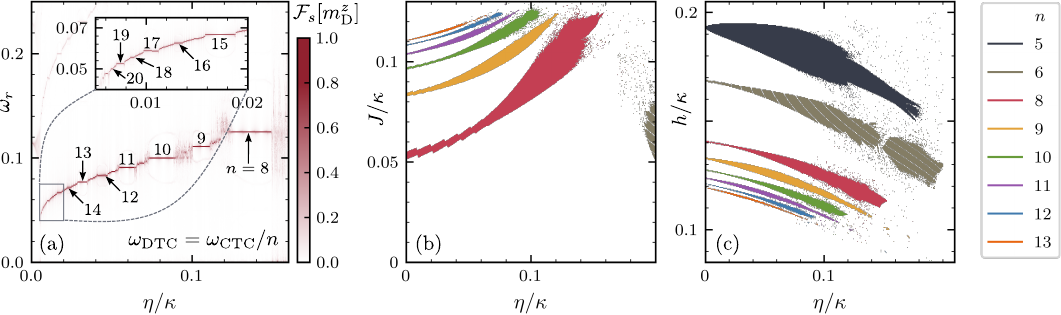}
        \caption{\textbf{Hierarchical symmetry breaking via spin-exchange coupling in the mean-field limit.} (a) Relative Fourier amplitudes of the DTC magnetization z component $\mathcal{F}_r[m_{\text{D}}^z]$ are plotted against the rescaled inter-TC coupling strength $\eta/\kappa$. The dominant component of the relative frequency $\omega_r$ exhibits distinct locking plateaus at both integer and fractional values for the DTC order $n$. System parameters are set to $\Omega/\kappa$ = 2, $J/\kappa$ = 0.1157, and $h/\kappa$ = 0.115, while the initial state is $(m^x_\text{C,D},m^y_\text{C,D},m^z_\text{C,D})=(0,0,1)$. The main panel and inset probe mz2 over 1000 and 3000 CTC periods, respectively. (b,c) Stability regimes under variation of the DTC all-to-all interaction strength $J$ and drive strength $h$, respectively. Dominant relative frequencies of the corresponding $n$-DTC are marked in their respective color.}
        \label{fig: Master figure Spin exchange coupling}
    \end{figure}
    
    To identify HTCs, we stick with the ratio of the subsystem oscillation frequencies $\omega_r$ as an order parameter, introduced in the main text and obtained via Fourier transformations of the mean-field solutions $m_i^{\alpha}(t)$. Displayed in Fig.~\ref{fig: Master figure Spin exchange coupling}(a), constant plateaus where the DTC frequency locks to a subharmonic frequency of the CTC emerge. 
    Interestingly, while for the considered coherent and dissipative coupling schemes in the main text, the order $n$ of the locking plateaus increased with growing coupling strength $\eta$, the opposite behavior is observed here.
    
    Notably, higher DTC orders, given by $n$, are observed compared to the discussed coupling schemes in the main text, reaching up to $n=20$. For lower DTC orders, we verify that the found stable HTC phases persist under variation of the remaining DTC parameters, namely the all-to-all interaction strength $J$ and the constant drive $h$. This can be seen in Fig.~\ref{fig: Master figure Spin exchange coupling}(b) and (c), where points corresponding to stable subharmonic frequency locking are marked, while the color indicates the DTC order $n$. Again, one finds stability areas with respect to different parameters, confirming the robust nature of the many-body HTC phase. 
    In conclusion, we have demonstrated that spin-exchange coupling also gives rise to hierarchical time translational symmetry breaking and, consequently, a stable HTC phase.

\section{Detailed derivation of the considered HTC models}
    We provide detailed microscopic derivations for both discussed time crystal models, as well as for the different considered coupling schemes between these subsystems. Although implementations based on Bose-Einstein condensates~\cite{KongkhmabutObservation2, Kongkhambut_CTC} or exciton-polariton condensates~\cite{autti2021ac, doi:10.1126/science.adn7087, PhysRevLett.120.215301, fainstein2026synchronization} are also possible, we focus here on cavity-QED platforms. We first focus on the CTC and DTC models and then turn to engineering their mutual interaction.
    \subsection{The boundary time crystal model}
        To realize the boundary time crystal model in Eq.~(2) of the main text, we follow Refs.~\cite{iemini2018btc, Seeding_crystallization} and consider an ensemble of $N_C$ identical two-level atoms with associated Pauli operators $\hat{\sigma}_{i,C}^{\alpha}$ in direction $\alpha\in\{x,y,z\}$ and $\hat{\sigma}_{i,C}^\pm = \hat{\sigma}_{i,C}^x \pm i\hat{\sigma}_{i,C}^y$. 
        The atoms are placed inside a lossy cavity which loses excitations to the environment at a rate $\gamma_C$ and is coherently driven with amplitude $\beta$. Assuming a Tavis-Cummings type interaction with strength $g_C$ between the atoms and the cavity, the dynamics of the system can be described by a Lindblad master equation of the form
        \begin{align}
        \label{supp: master equation microscopic CTC}
            \dot{\hat{\rho}} = -i[\hat{H}_\text{CTC}, \hat{\rho}] + \gamma_C \mathcal{D}[\hat{a}]\hat{\rho} \quad \text{where}\quad \hat{H}_\text{CTC} = g_C \sum_{i=1}^{N_C} (a^\dagger \hat{\sigma}_{i,C}^{-} + \hat{\sigma}_{i,C}^{+} a) - i\beta(a-a^\dagger)\;.
        \end{align}
        Upon introducing the collective spin operators as $\hat{S}_C^\alpha = \sum_{i=1}^{N_C} \hat{\sigma}_{i,C}^{\alpha}/2$, one finds that the Heisenberg equation of motion for the bosonic annihilation operator $\hat{a}$ reads
        \begin{align}
            \frac{d\hat{a}}{dt} = -2ig_C \hat{S}_C^- + \beta - \frac{\gamma_C}{2} \hat{a}\;.
        \end{align}
        In the bad cavity limit where $\gamma_C \gg g_C,\beta$, the cavity relaxes on a short timescale compared to the system evolution, allowing us to set its time derivative to zero.
        In this limit, one thus finds that
        \begin{align}
            \hat{a}(t) = -\frac{4ig_C}{\gamma_C} \hat{S}_C^- + \frac{2\beta}{\gamma_C} \;.
        \end{align}
        Substituting this result back into the master equation~(\ref{supp: master equation microscopic CTC}) yields
        \begin{align}
            \dot{\hat{\rho}} = \mathcal{L}_{\text{CTC}}\rho=-i\left[\frac{8\beta g_C}{\gamma_C} \hat{S}_C^x , \hat{\rho}\right] + \frac{16 g_C^2}{\gamma_C} \mathcal{D}[\hat{S}_C^-]\hat{\rho} \;.
        \end{align}
        This therefore exactly reproduces the master equation for the CTC used in Eq.~(2) of the main text when defining
        \begin{align}
            \Omega \equiv \frac{16g_C^2}{\gamma_C} \quad \text{and} \quad \kappa \equiv \frac{4\beta g_C N_C}{\gamma_C} \;.
        \end{align}

    \subsection{LMG model}
        In contrast to the CTC realization, which exploits the bad-cavity limit, the long-range interactions of the LMG model can be engineered in the dispersive regime of a far-detuned cavity~\cite{sauerwein2023engineering, Mivehvar02012021}.
        Analogous to the CTC case, we consider an ensemble of $N_D$ spin-$1/2$ systems with level splitting $\omega_a$, which interacts with a lossy cavity mode $\hat{b}$ at strength $g_D$. We denote the cavity frequency as $\omega_c$ and the cavity loss rate as $\gamma_D$. In the rotating frame of the atomic frequency $\omega_a$, the system Hamiltonian thus becomes
        \begin{align}
        \label{supp: microscopic hamiltonian LMG}
            \hat{H}_\text{DTC} = -\Delta_{ac} \hat{b}^\dagger \hat{b} + g_D \sum_{i=1}^{N_D}(\hat{b}^\dagger \hat{\sigma}_{i,D}^{-} + \hat{\sigma}_{i,D}^{+} \hat{b}) \;,
        \end{align}
        with $\Delta_{ac} = \omega_a -\omega_c$ the detuning between the cavity and the atomic transition. Also taking into account the dissipation of the cavity, the Heisenberg equation of motion for the cavity field thus takes the form
        \begin{align}
        \label{supp: Heisenberg eqm LMG model}
            \frac{d\hat{b}}{dt} = i\Delta_{ac} \hat{b} - 2ig_D \hat{S}_D^- -\frac{\gamma_D}{2}\hat{b} \;,
        \end{align}
        where we again used the collective spin operators introduced in the previous section.
        In the dispersive, good-cavity regime $\Delta_{ac} \gg \gamma_D$, the cavity losses in Eq.~(\ref{supp: Heisenberg eqm LMG model}) are negligible and the field adiabatically follows the spin dynamics, such that
        \begin{align}
            \hat{b}(t) = \frac{2g_D}{\Delta_{ac}}  \hat{S}_D^- \;.
        \end{align}
        Plugging this result back into the Hamiltonian in Eq.~(\ref{supp: microscopic hamiltonian LMG}) yields
        \begin{align}
        \label{supp: LMG Hamiltonian}
            \hat{H}_\text{DTC} = \frac{4g_D^2}{\Delta_{ac}} \hat{S}_D^+ \hat{S}_D^- = -\frac{4g_D^2}{\Delta_{ac}} \left((\hat{S}_D^z)^2 - (\hat{\Vec{S}}_D)^2 - \hat{S}_D^z \right) \;.
        \end{align}
        Since the total spin is conserved, the second contribution in Eq.~(\ref{supp: LMG Hamiltonian}) is constant and can be dropped, while the last term $\sim \hat{S}_D^z$ can be removed by an additional rotating-frame transformation.
        We are therefore left with the LMG Hamiltonian presented in Eq.~(3) of the main paper upon defining
        \begin{align}
            J \equiv \frac{g_D^2 N_D}{\Delta_{ac}} \;.
        \end{align}
        After discussing the standalone time crystal subsystems, we now turn to implementing the discussed coupling schemes. In general, this requires an additional cavity mode shared by both subsystems that mediates the interaction.
        
    \subsection{Dissipative coupling}
        In order to engineer collective dissipation of the time crystal subsystems, we consider a shared lossy cavity mode $\hat{e}$, which loses excitations to the environment at a rate $\gamma_I$. Both subsystems couple to this mode at strength $g_I$, described by the Hamiltonian 
        \begin{equation}
            \begin{aligned}
                \hat{H}_\text{Int} &= g_I \sum_{i=1}^{N_C} (\hat{e}^\dagger \hat{\sigma}_{i,C}^- + \hat{\sigma}_{i,C}^+ \hat{e}) + g_I \sum_{j=1}^{N_D} (\hat{e}^\dagger \hat{\sigma}_{j,D}^- + \hat{\sigma}_{j,D}^+ \hat{e}) \\
                &= 2g_I \left( \hat{e}^\dagger(\hat{S}_C^- + \hat{S}_D^-) + \hat{e}(\hat{S}_C^+ + \hat{S}_D^+)\right)\;.
            \end{aligned}
        \end{equation}
        Here, we again used the collective spin operators introduced in the previous sections. 
        The corresponding master equation thus reads
        \begin{align}
        \label{supp: master equ dissipative coupling microscoopic}
            \dot{\hat{\rho}} =\mathcal{L}_{\text{CTC}} \hat{\rho} + \mathcal{L}_{\text{DTC}} \hat{\rho} -i[\hat{H}_\text{Int}, \hat{\rho}] + \gamma_I \mathcal{D}[\hat{e}]\hat{\rho}\;,
        \end{align}
        which leads to the Heisenberg equation of motion for $\hat{e}$ of the form
        \begin{align}
            \frac{d\hat{e}}{dt} = -2ig_I (\hat{S}_C^- + \hat{S}_D^-) - \frac{\gamma_I}{2}\hat{e}\;.
        \end{align}
        In the bad cavity limit where $\gamma_I$ is the dominant energy scale, we set the time derivative of $\hat{e}$ to zero and obtain
        \begin{align}
            \hat{e}(t) = -\frac{4ig_I}{\gamma_I} (\hat{S}_C^- + \hat{S}_D^-)\;.
        \end{align}
        Substituting this result into the master equation~(\ref{supp: master equ dissipative coupling microscoopic}) yields
        \begin{align}
            \dot{\hat{\rho}} = \mathcal{L}_{\text{CTC}}\rho+\mathcal{L}_{\text{DTC}}\rho+\eta \mathcal{D}[\hat{S}_C^- + \hat{S}_D^-] \hat{\rho},
        \end{align}
        where the last term corresponds to the collective dissipation interaction scheme between the time crystal setups discussed in Eq.~(5) of the main text, with $\eta=16 g_I^2/\gamma_I^2$ being the collective dissipation strength.

    \subsection{Coherent coupling}
        The coherent coupling scheme introduced in Eq.~(4) of the main text can be engineered by coupling both time crystal subsystems to a shared cavity mode $\hat{e}$ in the dispersive regime at strengths $\lambda_C$ and $\lambda_D$, respectively. For the subsystem-cavity coupling, we assume Dicke-type interactions such that the interaction Hamiltonian in a chosen rotating frame takes the form
        \begin{align}
            H_\text{Int}^{(1)} = - \Delta_e \hat{e}^\dagger \hat{e} + \lambda_C (\hat{e}  + \hat{e}^\dagger) \hat{S}_C^y + \lambda_D(\hat{e}  + \hat{e}^\dagger) \hat{S}_D^x \;.
        \end{align}
        The respective Dicke interactions can be implemented using cavity-laser–mediated Raman transitions~\cite{DickeCouplingRealization, Mivehvar02012021}. Analogous to the preceding two sections, we can formulate the Heisenberg equation of motion for the cavity field. 
        By assuming that the cavity operates in the dispersive regime, we find that $\hat{e}(t) = (\lambda_C \hat{S}_C^y + \lambda_D \hat{S}_D^x)/\Delta_e$ and consequently
        \begin{align}
        \label{supp: yx coupling hint microcopic}
            \hat{H}_\text{int}^{(1)} = \frac{1}{\Delta_e} \left (2\lambda_C \lambda_D\hat{S}_C^y \hat{S}_D^x +\lambda_C^2 (\hat{S}_C^y)^2  +\lambda_D^2 (\hat{S}_D^x)^2\right)\;.
        \end{align}
        While the first term in Eq.~(\ref{supp: yx coupling hint microcopic}) already has the form of the desired coherent coupling Hamiltonian, we find additional local contributions for the individual subsystems.
        This can be resolved by considering a second shared cavity mode $f$ with opposite detuning $\Delta_f = -\Delta_e$ and Dicke-type coupling, such that the corresponding Hamiltonian reads 
        \begin{align}
            H_\text{Int}^{(2)} = - \Delta_f \hat{f}^\dagger \hat{f} + \lambda_C (\hat{f}  + \hat{f}^\dagger) \hat{S}_C^y - \lambda_D(\hat{f}  + \hat{f}^\dagger) \hat{S}_D^x \;.
        \end{align}
        Following the same steps as before, one finds that
        \begin{align}
            \hat{H}_\text{int}^{(2)} = \frac{1}{\Delta_f} \left (-2\lambda_C \lambda_D\hat{S}_C^y \hat{S}_D^x +\lambda_C^2 (\hat{S}_C^y)^2  +\lambda_D^2 (\hat{S}_D^x)^2\right)\;, 
        \end{align}
        and therefore finally
        \begin{align}
            \hat{H}_\text{int} = \hat{H}_\text{int}^{(1)} + \hat{H}_\text{int}^{(2)} = \frac{4 \lambda_C \lambda_D}{\Delta_e}\hat{S}_C^y \hat{S}_D^x\;.
        \end{align}
        By defining 
        \begin{align}
            \eta \equiv \frac{2\lambda_c \lambda_D N}{\Delta_e} \;,
        \end{align}
        we reproduce the coherent coupling scheme discussed in Eq.~(4) of the main text (see also Appendix~\ref{sec: coherent coupling}).

        Note that for the additional coherent spin-exchange coupling scheme introduced in Appendix~\ref{supp: sec spin exchange}, one proceeds analogously, but starting from a Tavis-Cummings-like interaction between the additional cavity mode $\hat{e}$ and the subsystems.

\end{document}